\documentclass[pra]{revtex4}
\usepackage{amsmath}
\usepackage{amssymb}
\usepackage{amsthm}
\usepackage{epsfig}
\usepackage{subfigure}
\newtheorem{theorem}{Theorem}

\DeclareMathOperator{\Tr}{Tr}
\begin{document}
\title{Minimax discrimination of two Pauli
  channels}
  \author{G. Mauro D'Ariano}\email{dariano@unipv.it}
    \author{Massimiliano F.
    Sacchi}\email{msacchi@unipv.it}\affiliation{{\em QUIT Group} of
    the INFM, Unit\`a di Pavia} \affiliation{Universit\`a di Pavia,
    Dipartimento di Fisica ``A. Volta'', via Bassi 6, I-27100 Pavia,
    Italy}
  \homepage{http://www.qubit.it} \author{Jonas
  Kahn}\email{jokahn@clipper.ens.fr} \affiliation{Universit\'e
  Paris-Sud 11, D\'epartement de Math\'ematiques, B\^at 425 91405
  Orsay Cedex (France)} 
\date{\today}
\begin{abstract} 
We consider the problem of optimally discriminating two Pauli channels
in the minimax strategy, maximizing the smallest of the probabilities
of correct identification of the channel. We find the optimal input
state at the channel and show the conditions under which using
entanglement strictly enhances distinguishability.  We finally
compare the minimax strategy with the Bayesian one.
\end{abstract}
\date{\today}
\pacs{}
\maketitle
\section{Introduction}
The concept of distinguishability applies to quantum states
\cite{distmeas1} and quantum processes \cite{distmeas2}, and is
strictly related to quantum nonorthogonality, a basic feature of
quantum mechanics. The problem of discriminating nonorthogonal quantum
states has been extensively addressed \cite{rev12}, also with
experimental demonstrations \cite{exper}. Typically, two
discrimination schemes are considered: the minimal-error probability
discrimination \cite{hel}, where each measurement outcome selects one
of the possible states and the error probability is minimized, and the
optimal unambiguous discrimination \cite{unam}, where unambiguity is
paid by the possibility of getting inconclusive results from the
measurement. The problem has been analyzed also in the presence of
multiple copies \cite{acin2}, and for bipartite quantum states, and
global joint measurements have been compared to LOCC measurements,
i.e. local measurements with classical communication
\cite{walg,shash,chin}. More recently, the discrimination of quantum
states has been addressed in the minimax approach \cite{freq}, where
there are no {\em a priori} probabilities, and one maximizes the
smallest of the probabilities of correct detection.  In such a scheme,
interesting results have been obtained, as, for example, optimal
solutions that involve unique and nonorthogonal measurements.

\par The problem of discrimination can be addressed also for quantum
operations \cite{discr}. This may be of interest in quantum error
correction \cite{ec}, since knowing which error model is the proper
one influences the choice of the coding strategy as well as the error
estimation employed. Clearly, when a repeated use of the quantum
operation is allowed, a full tomography can identify it. On the other
hand, a discrimination approach can be useful when a restricted number
of uses of the quantum operation is available.  Differently from the
case of discrimination of unitary transformations \cite{CPR}, for
quantum operations there is the possibility of improving the
discrimination by means of ancillary-assisted schemes such that
quantum entanglement can be exploited \cite{discr}.  Notably,
entanglement can enhance the distinguishability even for
entanglement-breaking channels \cite{entb}. The use of an arbitrary
maximally entangled state turns out to be always an optimal input when
we are asked to discriminate two quantum operations that generalize
the Pauli channel in any dimension. Moreover, in the case of Pauli
channels for qubits, a simple condition reveals if entanglement is
needed to achieve the ultimate minimal error probability
\cite{discr,pauli}. The above statements about channel discrimination
refer to a Bayesian approach.

\par In this paper we address the problem of optimal discrimination of
two Pauli channels in the minimax game-theoretical scenario.
Similarly to the case of state discrimination, we will show that the
two approaches generally give different results. In Sec. II we briefly
review the problem of discrimination of two Pauli channels in the
Bayesian framework, where the channels are supposed to be given with
assigned {\em a priori} probabilities. We report the result for the
optimal discrimination, along with the condition for which
entanglement with an ancillary system at the input of the channel
strictly enhances the distinguishability. In Sec. III we review the
solution to the problem of state discrimination in the minimax
approach, and its relation with the Bayesian problem. In Sec. IV we
study the problem of discrimination of two Pauli channels in the
minimax approach. We show that when an entangled-input strategy is
adopted, the optimal discrimination can always be achieved by sending
a maximally entangled state into the channel, as it happens in the
Bayesian approach.  On the contrary, the optimal input state for a
strategy where no ancillary system is used can be different in the
minimax approach with respect to the Bayesian one. In the latter the
optimal input can always be chosen as an eigenstate of one of the
Pauli matrices, whereas in the former this may not be the case. In the
concluding section, we summarize the main results of the paper.
\section{Bayesian discrimination of two Pauli channels}
\label{BDPC} 
In the problem of optimal Bayesian discrimination of two quantum
states $\rho _1$ and $\rho _2$, given with {\em a priori} probability
$p_1=p$ and $p_2=1-p$, respectively, one has to look for the
two-values probability operator-valued measure (POVM) $\vec B \equiv
\{B_1, B_2 \}$ with $B_i\geq 0$ for $i=1,2$ and $B_1+ B_2=I $, that
minimizes the error probability (or ``Bayes risk'')
\begin{eqnarray}
R_B(p, \vec B)=p_1 \hbox{Tr}[\rho _1 B_2] + p_2 \hbox{Tr}[\rho_ 2 B_1]\;. 
\end{eqnarray}
We can rewrite
\begin{eqnarray}
R_B(p, \vec B) &=& p_1 - \hbox{Tr}[(p_1 \rho _1 -p_2 \rho_2 ) B_1]
\nonumber \\&= & 
p_2 + \hbox{Tr}[(p_1 \rho _1 -p_2 \rho_2 ) B_2]
\nonumber \\&= & 
\frac 12 \left \{
1- \hbox{Tr}[(p_1 \rho _1 -p_2 \rho_2 ) (B_1 - B_2)]\right \}
\;, 
\end{eqnarray}
where the third line can be obtained by summing and dividing the two
lines above. The minimal error probability $R_B(p)\equiv \min_{\vec B}
R_B(p, \vec B)$ can then be achieved by taking the orthogonal POVM
made by the projectors on the support of the positive and negative
part of the Hermitian operator $p_1 \rho_ 1 -p_2 \rho _2$, and hence
one has \cite{hel,shash}
\begin{eqnarray}
R_B(p)= \frac 12 \left (1 -\Vert p_1 \rho_ 1 -p _2 \rho _2 \Vert _1
\right )\;,\label{pest}
\end{eqnarray}
where $\Vert A\Vert _1 =\hbox{Tr}\sqrt{A^\dag A} $ denotes the trace
norm of $A$. Notice that the optimal POVM does not appear in the
expression of the minimal error probability (\ref{pest}), as the trace
norm implicitly takes it into account.  \par The problem of optimally
discriminating two quantum operations ${\cal E}_1$ and ${\cal E}_2$
can be reformulated into the problem of finding the state $\rho $ in
the input Hilbert space $\cal H$, such that the error probability in
the discrimination of the output states ${\cal E}_1 (\rho )$ and
${\cal E}_2(\rho )$ is minimal.  The possibility of exploiting
entanglement with an ancillary system can increase the
distinguishability of the output states \cite{discr}. In this case the
output states to be discriminated will be of the form $({\cal
E}_1\otimes {\cal I}_{\cal K} ) \rho $ and $({\cal E}_2\otimes {\cal
I}_{\cal K}) \rho $, where the input $\rho $ is generally a bipartite
state of ${\cal H}\otimes {\cal K}$, and the quantum operations act
just on the first party whereas the identity map ${\cal I}_{\cal K}$
acts on the second.  \par Upon denoting with ${\cal R}' _B (p) $ the
minimal error probability when a strategy without ancilla is adopted,
one has
\begin{eqnarray}
{\cal R}'_B (p) 
=\frac 12 \left (1- \max _{\rho \in {\cal H}}\Vert p_1 {\cal E}_1
(\rho )- p_2{\cal E}_2(\rho )\Vert  _1\right )
\;.\label{peno}
\end{eqnarray}
On the other hand, by allowing the use an ancillary system, we have
\begin{eqnarray}
{\cal R}_B (p) =\frac 12 \left (1- \max _{\xi \in {\cal H}\otimes {\cal K}}
\Vert p_1 ({\cal E}_1 \otimes {\cal I})
\xi  - p_2 ({\cal E}_2\otimes {\cal I})\xi \Vert  _1\right )
\;.\label{pesi}
\end{eqnarray}
The maximum of the trace norm in Eq. (\ref{pesi}) with the supremum
over the dimension of ${\cal K}$ is equivalent to the norm of complete
boundedness \cite{paulsen} of the map $p_1 {\cal E}_1-p_2 {\cal E}_2$,
and in fact for finite-dimensional Hilbert space the supremum is
achieved for $\hbox{dim}({\cal K})=\hbox{dim}({\cal H})$
\cite{paulsen,diam}, and in the following we will drop the subindex
${\cal K}$ from the identity map. Moreover, due to linearity of
quantum operations and convexity of the trace norm, the maximum in
both Eqs. (\ref{peno}) and (\ref{pesi}) is achieved on pure states.
\par Clearly, ${\cal R}_B(p) \leq {\cal R}'_B(p) $. In the case of
discrimination between two unitary transformations $U$ and $V$
\cite{CPR}, one has ${\cal R}_B(p) = {\cal R}'_B(p)$, namely there is
no need of entanglement with an ancillary system to achieve the
ultimate minimum error probability, which is given by
\begin{eqnarray}
{\cal R}_B(p) &=&\min _{|\psi \rangle \in {\cal H}} \frac12 \left ( 1-
\sqrt{1-4p_1p_2 |\langle \psi |U^\dag V|\psi \rangle |^2}\right )
\nonumber \\& =& \frac12 \left ( 1- \sqrt{1-4p_1p_2 D^2}\right )
\;,\label{peuv}
\end{eqnarray}
where $D$ is the distance between $0$ and the polygon in the complex
plane whose vertices are the eigenvalues of $U^\dag V$.
\par In the case of discrimination of two Pauli channels for qubits, namely
\begin{eqnarray}
{\cal E}_i (\rho )= \sum_{\alpha =0}^3 q_\alpha ^{(i) }\sigma
  _\alpha \rho \sigma _\alpha \;\qquad i=1,2\;, \label{pc}
\end{eqnarray}
where $\sum _{\alpha =0}^3 q_\alpha ^{(i)} = 1$, $\sigma _0 = I$, and
$\{\sigma _1\,,\sigma _2\,,\sigma _3 \}= \{\sigma _x\,,\sigma
_y\,,\sigma _z\}$ denote the customary spin Pauli matrices, the
minimal error probability can be achieved by using a maximally
entangled input state, and one obtains \cite{discr}
\begin{eqnarray}
{\cal R}_B(p)= 
\frac 12 \left(1- \sum _{\alpha =0}^3 |r_\alpha | \right)
\;, \label{pe12}
\end{eqnarray}
with 
\begin{eqnarray}
r_\alpha = p_1 q_\alpha ^{(1)}- 
p_2 q_\alpha ^{(2)}=p(q_\alpha ^{(1)}+q_\alpha ^{(2)}) - q_\alpha ^{(2)}
\;,\label{ralpha}
\end{eqnarray}
where we fixed the {\em prior} $p=p_1$ and $p_2=1-p_1$.  For a
strategy with no ancillary assistance one has \cite{discr}
\begin{eqnarray}
{\cal R}'_B(p) = \frac 12 \left (1 - C \right ) \;, \label{pem}
\end{eqnarray}
where 
\begin{eqnarray}
C=\max \left\{ |r_0+ r_3|+|r_1+r_2|\,, |r_0+ r_1|+|r_2+r_3|\,, |r_0+
r_2|+|r_1+r_3| \right \} \;,\label{emme}
\end{eqnarray}
and the three cases inside the brackets corresponds to using an
eigenstate of $\sigma _z$, $\sigma _x$, and $\sigma _y$, respectively,
as the input state of the channel.  More generally, for pure input
state $\rho =\frac 12 (I + \vec \sigma \cdot \vec n )$, with $\vec
n=(\sin \theta \cos \phi, \sin\theta \sin \phi, \cos \theta )$, the
Bayes risk for discriminating the outputs will be \cite{discr,pauli}
\begin{eqnarray}
\label{bayesinput}
&& {\cal R}'_B (p,\vec \sigma \cdot \vec n) = \nonumber \\& &
\frac{1}{2} \left( 1 - \max \left\{ |a+b| , \sqrt{\cos ^2 \theta
(a-b)^2+ \sin ^2 \theta (c^2 + d^2 + 2cd \cos(2 \phi))} \right \}
\right)\;,
\end{eqnarray}
with $a = r_0 + r_3$, $b = r_1 + r_2$, $c = r_0 - r_3$, and $d = r_1 -
r_2$.  Notice that the term $|a+b |=|2p -1|$ corresponds to the
trivial guessing $\{ {\cal E}_1 \hbox{ if } p_1=p > 1/2\,, {\cal E}_2
\hbox{ if } p< 1/2 \}$.  \par\noindent We can also rewrite Eq.
(\ref{pem}) as
\begin{eqnarray}
{\cal R}'_B(p) = \min _{i=1,2,3}{\cal R}'_B(p, \sigma _i) \;.
\label{pem2}
\end{eqnarray}
\par From Eqs. (\ref{pe12}--\ref{emme}) one can see that entanglement
is not needed to achieve the minimal error probability as long as
$C=\sum _{i=0}^3 |r_i|$, which is equivalent to the condition $\Pi
_{i=0}^3 r_i \geq 0$. On the other hand, we can find instances where
the channels can be perfectly discriminated only by means of
entanglement, for example in the case of two channels of the form
\begin{eqnarray}
{\cal E}_1(\rho )=\sum _{\alpha \neq \beta }q_\alpha \sigma _\alpha
  \rho \sigma _\alpha \;,\qquad {\cal E}_2(\rho )=\sigma _\beta \rho
  \sigma _\beta \;,
\end{eqnarray}
with $q_\alpha \neq 0$, and arbitrary {\em a priori} probability. 
\section{Minimax discrimination of quantum states}
In the following we briefly review some results of Ref. \cite{freq}
about minimax discrimination of quantum states that are needed to
solve the problem of discrimination of Pauli channels in the next
Section, namely we review just the case of two states.  We are given
two states $\rho_1$ and $\rho_2$, and we want to find the optimal
measurement to discriminate between them in a minimax approach. In
this scenario there are no {\em a priori} probabilities, and the
optimal solution consists in finding the POVM $\{\vec M = M_1, M_2\}$
with $M_i\geq 0$ for $i=1,2$ and $M_1+ M_2=I$, that achieves the
minimax
\begin{equation}
R_M (\rho _1, \rho _2) 
=\min_{\vec M}
\max \{ \Tr [\rho _1 M_2], \Tr [\rho _2 M_1]\}\;,\label{14}
\end{equation}
namely one minimizes the largest of the probabilities of incorrect
detection. The minimax and Bayesian schemes of discrimination of two
states are connected by the following theorems \cite{freq}
\begin{theorem}\label{l:bayes}
There is a measurement $\vec B$ that is optimal in the Bayes scheme
  for some {\em a priori} probability $(p_*,1-p_*)$ such that
\begin{equation}
\Tr[\rho_1 B_1]=\Tr[\rho_2 B_2]\;.\label{equalB}
\end{equation}
This measurement is optimal in the minimax scheme as well, and
one has $R_M(\rho _1, \rho _2) = R_B (p_*)= \Tr [\rho _1 B_2]$.  
\end{theorem}
\begin{theorem}\label{l:bayes2}
  The solution in the minimax problem is equivalent to the solution of
  the problem
\begin{eqnarray}
R_M(\rho _1, \rho _2)= \max _{p} R_B (p)\;,\label{maxpp}
\end{eqnarray}
and the {\em a priori} probability achieving the maximum corresponds
to the value $p=p_*$ in Theorem \ref{l:bayes}.
\end{theorem}
\section{Minimax discrimination of Pauli channels}
As in the Bayesian approach, the minimax discrimination of two
channels consists in finding the optimal input state such that the two
possible output states are discriminated with minimum risk. Again, we
will consider the two cases with and without ancilla, upon defining
\begin{eqnarray}
&&{\cal R}_M =\min _{\xi \in {\cal H}\otimes {\cal K}} R_M (({\cal
E}_1 \otimes {\cal I})(\xi) ,({\cal E}_ 2 \otimes {\cal I})(\xi )
)\;,  \nonumber \\& & {\cal R}_M '=\min _{\rho \in {\cal H}}
R_M ({\cal E}_1 (\rho ) ,{\cal E}_ 2 (\rho ))\;,  
\end{eqnarray}
where $R_M(\rho _1, \rho _2)$ is given in Eq. (\ref{14}).  
Since for all $\vec M$, $\rho $,  and $p$,  one has
\begin{eqnarray}
&&\max \{\Tr [({\cal E}_1\otimes {\cal I}) (\rho ) M_2] , \Tr [({\cal
E}_2 \otimes {\cal I})(\rho )M_1 ] \} \nonumber \\& & \geq p \Tr
[({\cal E}_1\otimes {\cal I}) (\rho ) M_2] + (1-p) \Tr [({\cal E}_2
\otimes {\cal I})(\rho )M_1 ] \;, \; 
\end{eqnarray}
then ${\cal R}_M \geq {\cal R}_B (p)$ for all $p$.  Analogously,
${\cal R}'_M \geq {\cal R}'_B (p)$ for all $p$.  

\par Theorems \ref{l:bayes} and \ref{l:bayes2} can be immediately
applied to state that the minimax discrimination of two unitaries is
equivalent to the Bayesian one. In fact, the optimal input state in
the Bayesian problem which achieves the minimum error probability of
Eq. (\ref{peuv}) does not depend on the {\em a priori} probabilities.
Therefore it is also optimal for the minimax problem and there is no
need of entanglement [and the minimax risk ${\cal R }_M$ will be
equivalent to the Bayes risk ${\cal R}_B (1/2)$].  \par Let us now
consider the problem of discriminating the Pauli channels of
Eq. (\ref{pc}) in the minimax framework. In the following theorem, we
show that an (arbitrary) maximally entangled state always allows to
achieve the optimal minimax discrimination as in the Bayesian problem.
\begin{theorem}\label{thment}
The minimax risk ${\cal R}_M$ for the discrimination of two Pauli
channels can be achieved by using an arbitrary maximally entangled
input state.  Moreover, the minimax risk is then the Bayes risk for
the worst \emph{a priori} probability:
\begin{equation}
{\cal R}_M= \max _{p} {\cal R}_B(p)\;.\label{supp}
\end{equation}
\end{theorem}
\begin{proof}
Let us discriminate between the states $\rho _i = (\mathcal{E}_i
\otimes {\cal I})(\xi ^e)$, where $\xi ^e$ is a maximally entangled
state. By Theorem \ref{l:bayes} there are \emph{a priori}
probabilities $(p_*,1-p_*)$ whose optimal Bayes measurement fulfills
\begin{equation} 
\Tr[\rho_1 B_1]=\Tr[\rho_2 B_2]\;.
\end{equation} 
Since the input state $\xi ^e$ is always optimal in the Bayes problem
we infer ${\cal R}_B(p_*)=\Tr[\rho_1 B_2] $, and moreover $R_M(\rho
_1, \rho _2) = {\cal R}_B(p_*)$. Now, one has also ${\cal R}_M =
R_M(\rho _1, \rho _2)$, since if it would not be true, then there
would be an input state $\rho $ and a measurement $\vec{M}$ for which
$\max \{\Tr[(\mathcal{E}_1\otimes {\cal I})(\rho) M_2],
\Tr[(\mathcal{E}_2\otimes {\cal I})(\rho) M_1]\} < R_B(p_*)$, and
hence $p_* \Tr[(\mathcal{E}_1\otimes {\cal I})(\rho) M_2] + (1-p_*)
\Tr[(\mathcal{E}_2\otimes {\cal I})(\rho) M_1] < R_B(p_*)$, which is a
contradiction.  Equation (\ref{supp}) simply comes from the relation
${\cal R}_M \geq {\cal R}_B (p)$ for all $p$, along with ${\cal R}_M
={\cal R}_B (p_*)$. 
\end{proof}
Notice the nice correspondence between Eqs.  (\ref{maxpp}) and
(\ref{supp}). Theorem \ref{thment} holds true also in the case of
generalized Pauli channels in higher dimension, since entangled states
again achieve the optimal Bayesian discrimination, whatever the
\emph{a priori} probability \cite{discr}. More generally,
Eq. (\ref{supp}) will hold in the discrimination of any couple of
quantum operations for which the minimal Bayes risk ${\cal R}_B (p)$
can be achieved by the same input state for any $p$.
\begin{figure}[ht]
\centering
\epsfig{figure=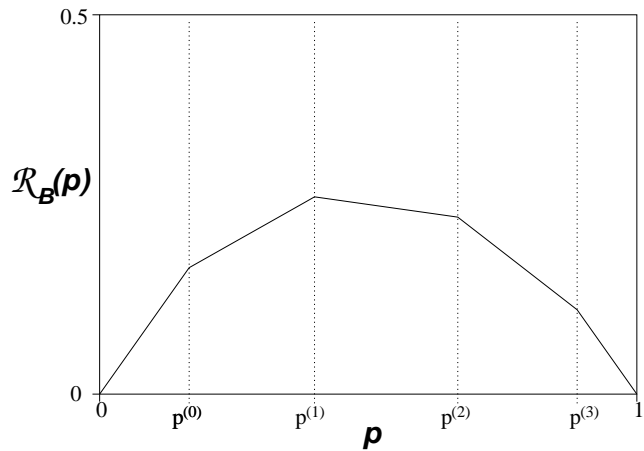,height=0.25\textheight}
\caption{\label{RbP} The optimal Bayes risk ${\cal R}_B(p)$ in the
  discrimination of two Pauli channels versus the {\em a priori}
  probability $p$ will usually look like this. Notice that the
  rightmost and leftmost segments have slope $1$ and $(-1)$,
  respectively. The minimal risk for the minimax discrimination
  corresponds to ${\cal R}_M =\max _p {\cal R}_B(p)$, and is achieved
  at one of the breakpoints $p^{(\alpha )}$.}
\end{figure}
\par Now we establish some visual images on which to read the minimax
risks. We must look at the function ${\cal R}_B(p)$ given in
Eq. (\ref{pe12}) drawn on $[0,1]$.  By Eq. (\ref{supp}), we know that
its maximum is ${\cal R}_M$. As the $r_{\alpha}$ defined in
(\ref{ralpha}) are increasing affine functions of $p$, their absolute
value is a convex piecewise affine function, and hence ${\cal R}_B(p)$
is a concave piecewise affine function (see Fig. \ref{RbP}). The four
breakpoints correspond to the four values of $p$ for which each
$r_{\alpha}$ vanishes.  We define $t_{\alpha}= q_{\alpha}^{(1)} +
q_{\alpha}^{(2)}$ as the slopes of the functions $r_{\alpha}$ and
$p^{(\alpha)}= q_{\alpha}^{(2)}/\,t_{\alpha}$ as the value of $p$ for
which $r_{\alpha}=0$. We denote by $p_*$ the point at which ${\cal
R}_{B}(p)$ reaches its maximum (the maximum will be attained at one of
the breakpoints $p^{(\alpha )}$). We also reorder the index $\alpha $
such that $p^{(0)} \leq p^{(1)} \leq p^{(2)} \leq p^{(3)}$. In this
way, ${\cal R}_B (p)$ rewrites
\begin{eqnarray}
{\cal R}_B (p)=\frac 12 \left
(1 -\sum _{\alpha =0}^3 t_\alpha  |p -p^{(\alpha )}|\right )\;. 
\end{eqnarray}
\par Let us now look at the discrimination strategy without any
ancillary system.  Another picture, that should be superimposed on
Fig. 1, is the Bayes risk ${\cal R}'_B (p)$ of Eq. (\ref{pem}) versus
$p$ for the strategy with no ancillary system.  One can see that
${\cal R}'_B(p)$ is the minimum of the three piecewise affine
functions ${\cal R}_B'(p, \sigma_x)$, ${\cal R}'_B(p, \sigma_y)$,
${\cal R}'_B(p, \sigma_z)$, corresponding to the Bayes risks when
sending an eigenstate of the Pauli matrices. Here again ${\cal
R}'_B(p)$ is the minimum of concave functions, so it is concave as
well, and the maximum will be attained at a breakpoint $p=p'_*$ (see
Fig. 2).
\begin{figure}[ht]
\centering
\epsfig{figure=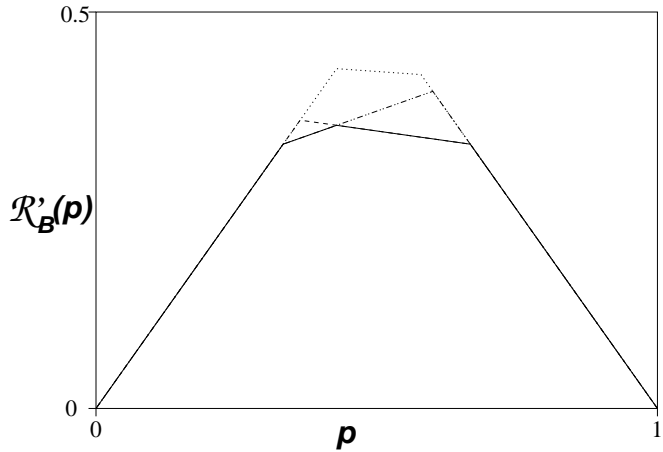,height=0.3\textheight}
\caption{\label{unent} An example for the Bayes risks ${\cal R}'_B (p,
  \sigma _i)$ with $i=x,y,z$ versus the {\em a priori} probability
  $p$, for discrimination without ancilla.  Each of the three
  different dotted lines correspond to the Bayes risk ${\cal R}_B'(p,
  \sigma _i)$ when sending an eigenstate of the Pauli matrix $\sigma
  _i$ through the channel. The solid line is the optimal Bayes risk
  ${\cal R}'_B(p)$ without ancillary assistance, and corresponds at
  any $p$ to the minimum of the three ${\cal R}'_B (p, \sigma
  _i)$. The minimal risk for the minimax discrimination with no
  ancilla corresponds to ${\cal R}'_M =\max _p {\cal R}'_B(p)$, and is
  achieved at one of the breakpoints of ${\cal R}'_B(p)$.}
\end{figure}
To ``read'' more on these pictures, once again we prove that the
optimal minimax risk ${\cal R}'_M$ for discrimination without ancilla
corresponds to the optimal Bayes risk without ancilla for the worst
\emph{a priori} probability $p_*'$:
\begin{theorem}
\label{unentangled}
The optimal minimax discrimination with no ancilla is equivalent
to the solution of the problem
\begin{equation}
\label{cu}
{\cal R}_M' = \max_{p} {\cal R}'_B(p) \equiv {\cal R}_B'(p_*') \;.
\end{equation} 
\end{theorem}
\begin{proof}
Notice again the similarity between equations (\ref{maxpp}),
(\ref{supp}) and (\ref{cu}). For any $\rho$ one has
\begin{eqnarray}
R_M ({\cal E}_1(\rho ), {\cal E}_2(\rho )) \geq {\cal R}_M' \geq \max
_p {\cal R}'_B (p) \;.\label{anyrho}
\end{eqnarray}
If we find an input state $\rho _{\vec n}=\frac 12 (I + \vec \sigma
\cdot \vec n)$ such that
\begin{eqnarray}
\max _p {\cal R}'_B (p) = \max _p {\cal R}'_B (p, \vec \sigma \cdot
\vec n)\;\label{etape}
\end{eqnarray}
from Eq. (\ref{maxpp}) of Theorem \ref{l:bayes2} it follows that 
\begin{eqnarray}
R_M ({\cal E}_1 (\rho _{\vec n}), {\cal E}_2 (\rho _{\vec n}))=
\max _p {\cal R}'_B (p, \vec \sigma  \cdot
\vec n)\;, 
\end{eqnarray}
which, along with Eqs. (\ref{anyrho}) and (\ref{etape}), provides the
proof. Moreover, $\rho _{\vec n}$ will be the optimal input state for
the minimax discrimination without ancilla.  \par\noindent Now we have
just to find a state such that condition (\ref{etape}) holds. We
already noticed that $p_*'$ is a breaking point of ${\cal R}'_B(p)$.
Either this breakpoint is also a breakpoint (and the maximum) of
${\cal R}'_B (p,\sigma_i)$ for some $i \in {x,y,z}$, or else at least
two of the ${\cal R}'_B (p,\sigma_i)$ are crossing in $p_*'$, one
increasing and the other decreasing (Fig. 2). In the first case
Eq. (\ref{etape}) is immediately satisfied, and an eigenstate of
$\sigma _i$ will be the optimal input state.  In the second case, we
show that when two ${\cal R}'_B (p,\sigma_i)$ are crossing at $p_*'$
we can find a state $\rho _{\vec n}$ such that
\begin{eqnarray}
&&{\cal R}_B'(p_*', \vec{\sigma}\cdot \vec n) = {\cal
R}'_B (p_*',\sigma_i) \;,\nonumber \\& & 
\partial _p {\cal R}_B'(p, \vec{\sigma}\cdot \vec{n} ) |_{p=p_*'}= 0
\;,\label{part}
\end{eqnarray}
and therefore has the maximum at $p_*'$ by concavity.  In fact, the
crossing, and therefore non-equality of the ${\cal R}'_B (p,\sigma_i)$
in a neighborhood of $p_*'$, implies that for each of the two ${\cal
  R}'_B(p, \sigma_i)$, the maximum in (\ref{bayesinput}) for $p_*'$ is
attained by the square root term (since the term $|a+b|$ is just a
function of $p$).  Let us assume that the $\sigma_i$ that give such a
crossing are $\sigma_x$ and $\sigma_y$. Then looking at
(\ref{bayesinput}), we have at point $p_*'$
\begin{eqnarray}
&&|c+d|=|c-d|\;,\nonumber \\
&&\partial _p |c+d| \, \partial _p |c-d| <0\; 
\label{28}
\end{eqnarray}
(notice that all functions are linear, i.e. differentiable in $p_*'$).
Indeed, the first of Eqs. (\ref{28}) implies that any linear
combination of eigenstate of $\sigma _x$ and $\sigma _y$ satisfies the
first of Eqs. (\ref{part}).  By taking an input state with $\theta
=\pi/2$ and $\phi $ such that
\begin{eqnarray}
\tan ^2 \phi= - \left.
\frac {\partial _p |c+d|}{\partial _p |c-d|}\right|_{p=p_*'}\;,\label{tanphi}
\end{eqnarray}
the second equation in (\ref{part}) is satisfied as well. Similarly,
if the $\sigma _i$ are $\sigma _z, \sigma _x$ one can take the input
state with $\phi =0$ or $\pi $ and $\theta $ such that
\begin{eqnarray}
  \tan ^2 \theta = - \left.
\frac {\partial _p |a-b|}{\partial _p |c+d|}\right|_{p=p_*'}\;.
\end{eqnarray}
Finally, for $\sigma _z, \sigma _y$ one has $\phi = \pm \pi/2$ and 
\begin{eqnarray}
\tan ^2 \theta= - \left.
\frac {\partial _p |a-b|}{\partial _p |c-d|}\right|_{p=p_*'}\;.
\; 
\end{eqnarray}
\par As a remark, no eigenstate of $\sigma_i$ for $i = x,y,z$ can be
an optimal input in the minimax sense in this situation.  This is a
typical result of the minimax discrimination.  As in the case of
discrimination of states \cite{freq}, when the correspondent Bayes
problem presents a kind of degeneracy and have multiple solutions, in
the minimax problem the degeneracy is partially or totally removed. In
the present situation, if we have the maximum of ${\cal R}_B'(p)$ at
the crossing point of exactly two ${\cal R}_B'(p, \sigma _i)$, one
increasing and the other decreasing, we find just four optimal input
states: two non-orthogonal states and their respective orthogonal
states. We will give an explicit example at the end of the section.
\end{proof}
If we want to find in which case entanglement is not necessary for
optimal minimax discrimination, then we have just to characterize when
${\cal R}_B'(p_*')={\cal R}_B(p_*)$.  We already noticed that we can
choose $p_*$ to be one of the $p^{(\alpha)}$. The corresponding
$r_{\alpha}$ is zero, and hence $C=\sum_\alpha |r_{\alpha}|$, namely $
{\cal R}_B'(p_*)={\cal R}_B(p_*)$. Since one has
\begin{eqnarray}
{\cal R}_B'(p_*') ={\cal R}_M'
\geq {\cal R}_M= {\cal R}_B(p_*) ={\cal
  R}_B' (p_*)
\;, 
\end{eqnarray}
we only have to check that $p_*$ is a maximum of $ {\cal R}_B'(p)$,
recalling that the function is concave (see
Fig. 3).
\begin{figure}[ht]
\centering
{\epsfig{figure=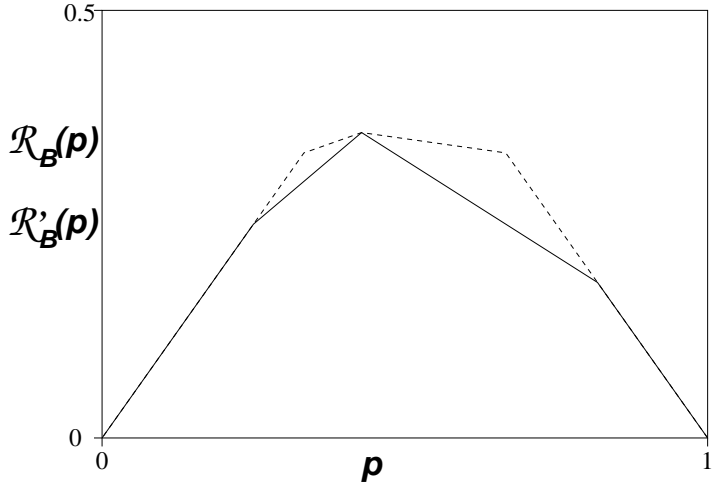,width=0.6\textwidth}\label{o2mmarche}}
\caption{Optimal Bayes risks versus the {\em a priori} probability $p$
  for the discrimination of the Pauli channels with parameters given  
  in Eq. (\ref{3cc}). The solid line gives ${\cal R}_B(p)$ for an
  entanglement-assisted strategy; the dotted lines gives ${\cal
  R}'_B(p)$ for strategy without ancilla. The minimal risk in the
  optimal minimax discrimination corresponds in both strategies
  to ${\cal R}'_M =\max _p {\cal R}'_B(p)=\max _p {\cal R}_B(p)={\cal
  R}_M$, namely there is no need of an ancillary system.}
\end{figure}
\par Ultimately, we will have to list down cases. Reading them might
be clearer with the quantities appearing in Eqs.
(\ref{pe12}--\ref{emme}) explicitly written as a function of $p$. The
most useful segmentation of $[0,1]$ is based on the $p^{(\alpha)}$,
that is the points where the $r_\alpha$ vanish, and ${\cal R}_B(p)$
breaks. Recall that $r_\alpha =t_\alpha (p-p^{(\alpha )})$, and
$r_\alpha >0$ for $p>p^{(\alpha )}$.  As we have four $\alpha$, we
have five segments (they may get degenerated).  Remember that knowing
$C$ in Eq. (\ref{emme}) and $\sum_\alpha |r_\alpha |$ is tantamount to
knowing $ {\cal R}'_B(p)$ or ${\cal R}_B(p)$. Here is a list of the
signs of the $r_\alpha$ and the value of $C$ on each open segment (so
that all $r_\alpha \neq 0$):
\begin{itemize}
\item{$(0,p^{(0)})$: $\sum_\alpha |r_\alpha|= - \sum_\alpha r_\alpha =
    C$. Notice that ${\cal R}_B' (p)= {\cal R}_B(p)$ and that their
    common slope is $1$.}
\item{$(p^{(0)},p^{(1)})$: $\sum_\alpha |r_\alpha| = r_0 - r_1 - r_2 -
    r_3$, so that $C = r_0 - r_1 - r_2 - r_3 - 2 \inf_{\alpha =1,2,3}
    |r_\alpha|$. On this segment, ${\cal R}_B' (p)> {\cal R}_B(p)$.}
\item{$(p^{(1)},p^{(2)})$ : $\sum_\alpha |r_\alpha| = r_0 + r_1 - r_2
    - r_3 = C$, so that ${\cal R}_B' (p)= {\cal R}_B(p)$.}
\item{$(p^{(2)},p^{(3)})$: $\sum_\alpha |r_\alpha| = r_0 + r_1 + r_2 -
    r_3$, so that $C = r_0 + r_1 + r_2 - r_3 - 2 \inf_{\alpha =0,1,2}
    r_\alpha $ and ${\cal R}_B' (p)> {\cal R}_B (p)$.}
\item{$(p^{(3)},1)$: $\sum_\alpha |r_\alpha|= \sum_\alpha r_\alpha =
    C$ and ${\cal R}_B' (p)= {\cal R}_B(p)$. Their common slope is
    $(-1)$.}
\end{itemize}
A close look at these expressions, as we will show in the following,
proves that ${\cal R}_B'(p)$ is derivable at $p^{(\alpha)}$ unless
there is $\beta \neq \alpha$ such that $p^{(\alpha)}=p^{(\beta)}$.
With this in mind, we see that $p_*$ cannot be a maximum of
$p^{(\alpha)}$ unless several $r_{\alpha}$ are null at the same point
(with supplementary conditions) or $p_* = p^{(1)}$ and the segment $
(p^{(1)},p^{(2)})$ is flat. Here is the full-fledged study, using
repeatedly the list above. It is complete as any other case can be
handled by symmetry (switching channels, that is mapping $p$ to
$1-p$).
\begin{itemize}
\item{$p_* = p^{(0)} < p^{(1)}$: At $p^{(0)}$, we have $r_0 = 0$ and
$r_\alpha < 0$ for $\alpha \neq 0$. So that $\inf_{\alpha}
|r_{\alpha}| = |r_0|$ on a neighborhood of $p^{(0)}$. On that
neighborhood, we deduce $C= - \sum_{\alpha} r_{\alpha}$, and hence
$\partial _p {\cal R}_B'(p) \vert _{p=p^{(0)}} =1$, so that $p^{(0)}$
is not a maximum of ${\cal R}_B'(p)$. Entanglement is then necessary
for optimal discrimination.}
\item{$p_* = p^{(0)} = p^{(1)} < p^{(2)}$: On $(0,p^{(0)}) \cup
    (p^{(1)},p^{(2)})$, equality ${\cal R}_B'(p)={\cal R}_B(p)$ holds.
    Thus, the two functions are equal on a neighborhood of $p_*$, and
    since $p_*$ is a (local) maximum of ${\cal R}_B(p)$, it is also a
    local maximum of ${\cal R}_B' (p)$. In this case an unentangled
strategy is then as efficient as any entangled one.}
\item{$p_* = p^{(0)} = p^{(1)} = p^{(2)} < p^{(3)}$: The risk ${\cal
      R}_B'(p)$ is nondecreasing on the left of $p_*$ (slope 1), we
    then want it to be non-increasing on a right neighborhood of
    $p_*$. Now this is part of the segment $(p^{(2)},p^{(3)})$, where
    $C = r_0 + r_1 + r_2 - r_3 -2 \inf_{\alpha =0,1,2} r_\alpha $.
    Recall that $r_\alpha =t_\alpha (p-p^{(\alpha )})$. Since
    $r_\alpha =0$ for $\alpha \neq 3$ at $p_*$, and they are all
    nondecreasing, $\inf_{\alpha =0,1,2} r_\alpha $ is the one with
    the smallest slope $t_\alpha $. It follows that the slope of
    ${\cal R}_B'(p)$ on the right of $p_*$ is $t_3 - t_0 - t_1 - t_2 +
    2 \inf_{\alpha =0,1,2} t_\alpha $, and so entanglement is not
    needed if and only if
\begin{equation}
\label{3p1}
  t_3 + 2 \inf_{\alpha ={0,1,2}} t_\alpha \leq \sum_{\alpha =0,1,2}
  t_\alpha
\end{equation}}
\item{$p_* = p^{(0)} = p^{(1)} = p^{(2)} = p^{(3)}$: This is the
    trivial case where both channels are the same.  Of course,
    entanglement is useless.}
\item{$p^{(0)} < p_* = p^{(1)} < p^{(2)}$: In this case ${\cal
      R}_B'(p)$ is derivable at $p_*$. Indeed, on $(p^{(1)},p^{(2)})$,
    we have $C= r_0 + r_1 - r_2 -r_3$ whereas on $(p^{(0)},p^{(1)})$,
    $ C = r_0 - r_1 - r_2 - r_3 - 2 \inf_{\alpha =1,2,3}
    |r_{\alpha}|$. In a neighborhood of $p_*$, one has $\inf_{\alpha
      =1,2,3} |r_{\alpha}|=r_1$, as it is the only one which is $0$ at
    $p_*$; hence $C= r_0 + r_1 - r_2 -r_3$ also on a left neighborhood
    of $p_*$ and the slope of ${\cal R}_B'(p)$ at $p_*$ is $t_3 + t_2
    - t_1 - t_0$. Since $p*$ is a maximum if and only if this slope is
    null, we get the condition
\begin{equation}
\label{1p1}
t_0 + t_1 = t_2 + t_3\;.
\end{equation}}
\item{$p^{(0)} < p_* = p^{(1)} = p^{(2)} < p^{(3)}$: On the left of
    $p_*$, we are on the segment $(p^{(0)},p^{(1)})$, so that $C= r_0
    -r_1 - r_2 - r_3 - 2 \inf_{\alpha=1,2,3} |r_{\alpha}|$. On the
    right, we are on the segment $(p^{(2)},p^{(3)})$ and $C= r_0 + r_1
    + r_2 - r_3 - 2 \inf_{\alpha=0,1,2} r_{\alpha}$. In a neighborhood
    of $p_*$, the $r_\alpha $ with the smallest absolute value will be
    either $r_1$ or $r_2$ (more precisely, the one with the smallest
    slope $t_\alpha $), so that we can write in a neighborhood of
    $p_*$ for both sides $C = r_0 - r_3 + |r_2 - r_1|$. The slope of
    ${\cal R}_B'(p)$ is then $t_3 - t_0 + |t_2 - t_1|$ and $ t_3 - t_0
    - |t_2 - t_1| $ on the left and on the right of $p_*$,
    respectively.  Entanglement is not necessary when $p_*$ is a
    maximum of ${\cal R}_B'(p)$, and hence we get the necessary and
    sufficient condition
\begin{equation}
\label{2p1}
 \left | t_0 - t_3 \right |  \leq \left | t_1 - t_2 \right |\;.
\end{equation}
}
\end{itemize}
We can summarize the above discussion as follows
\begin{theorem}
  The minimax risk without using ancilla is strictly greater than the
  minimax risk using entanglement, except in the following cases:
\begin{itemize}
\item{the trivial situation where both channels are the same, so that
    $p_*=p^{(\alpha )}=\frac 12$ for all $\alpha $.}
\item{if $p_*=p^{(0)} \leq p^{(1)}< p^{(2)}$}
\item{if $p_*=p^{(0)}=p^{(1)}= p^{(2)} < p^{(3)}$ and
\begin{equation}
  t_3 + 2 \inf_{\alpha =0,1,2} t_\alpha  \leq \sum_{\alpha =0,1,2}
  t_\alpha 
\end{equation}
}
\item{if $p^{(0)} < p_* = p^{(1)} < p^{(2)}$ and 
\begin{equation}
t_0 + t_1 = t_2 + t_3
\end{equation}}
\item{if $p^{(0)} < p_* = p^{(1)} = p^{(2)} < p^{(3)}$ and 
\begin{equation}
 \left| t_0 - t_3 \right| \leq \left| t_1 - t_2 \right|
\label{exem}
\end{equation}}
\item{The symmetric cases (obtained by exchanging channels 1 and 2,
    i.e. exchanging indexes 0 and 1 with 3 and 2, respectively, both
    in $p^{(\alpha )}$ and $t_\alpha $.}
\end{itemize} 
\end{theorem}
\par Differently from the Bayesian result, we notice that when
entanglement is not necessary to achieve the optimal minimax
discrimination, the optimal input state may not be an eigenstate of
the Pauli matrices. Consider, for example, the two Pauli channels
featured in Fig. 3 that correspond to the parameters
\begin{align}
q_0^{(1)} & = 0.3 & q_1^{(1)} & = 0.4 & q_2^{(1)} & = 0.2 & q_3^{(1)}
& = 0.1 \nonumber \\ 
q_0^{(2)} & = 0.1 & q_1^{(2)} & = 0.3 & q_2^{(2)} & = 0.15 & 
q_3^{(2)} & = 0.45 \label{3cc} 
\end{align}
We can compute $p^{(\alpha )}= q_\alpha ^{(2)}/(q_\alpha
^{(1)}+q_\alpha ^{(2)}) $ and get $p^{(\alpha )}=(1/4, 3/7,3/7,9/11)$.
Here $p_*=3/7$, and we are in the situation of Eq. (\ref{exem}), since
$t_\alpha =(q_\alpha ^{(1)}+q_\alpha ^{(2)})=(0.4,0.7,0.35,0.55)$.
Hence, entanglement is not necessary to achieve the optimal minimax
risk, but the state to be used is not an eigenstate of the Pauli
matrices. In fact, we are in the case of the proof of Theorem 3, where
${\cal R}_B'(p,\sigma _x)$ and ${\cal R}_B'(p,\sigma _y)$ are crossing
in $p_*$. The optimal input state for the minimax discrimination will
be given by $\theta =\pi/2$ and $\phi $ as in Eq. (\ref{tanphi}),
which gives $\tan ^2 \phi = 2/5$. Then, we have four optimal input
states that lie on the equator of the Bloch sphere, with $\vec n =(\pm
\sqrt{5/7},\pm \sqrt{2/7},0)$.
\section{Conclusions}
We addressed the problem of optimally discriminating two Pauli
channels in the minimax approach, where no {\em a priori} probability
is assigned.  We showed that when an entangled-input strategy is
adopted, the optimal discrimination can always be achieved by sending
a maximally entangled state into the channel, as it happens in the
Bayesian approach.  On the other hand, the optimal input state for a
strategy without ancilla can be different in the minimax approach with
respect to the Bayesian one. In the latter the optimal input can
always be chosen as an eigenstate of one of the Pauli matrices,
whereas in the former this may not be the case.  We then characterized
the channels where the use of entanglement outperforms the scheme
without assistance of ancilla.  Notice that even though the Bayesian
and the minimax strategies are not comparable, since they address
different estimation problems, nevertheless the solution of the
general Bayesian problem actually includes also the minimax solution,
since the optimal minimax strategy is equivalent to the Bayesian one
for the worst risk (Theorems 3 and 4). This is a general feature for
the channels analyzed in the present paper. This work extends the
study of minimax discrimination of states to the simplest example of
quantum operations, and show the relation and the differences with
respect to the Bayesian approach.
\section*{Acknowledgments } 
Support from INFM through the project PRA-2002-CLON, and from EC and
MIUR through the cosponsored ATESIT project IST-2000-29681 and
Cofinanziamento 2003 is acknowledged.

\end{document}